\documentclass[twocolumn,english,pra,showpacs]{revtex4-1}
\usepackage[T1]{fontenc}
\usepackage{babel}
\usepackage{amsmath}
\usepackage{amssymb}
\usepackage{graphicx}
\usepackage{esint}
\usepackage[unicode=true,pdfusetitle,
 bookmarks=true,bookmarksnumbered=false,bookmarksopen=false,
 breaklinks=false,pdfborder={0 0 1},backref=section,colorlinks=false]
 {hyperref}
\hypersetup{
 colorlinks,linkcolor=blue,citecolor=blue}

\usepackage{babel}

\begin{document}

\title{Entanglement Dynamics of Two Qubits in a Common Bath}

\author{Jian Ma$^{1,2}$, Zhe Sun$^{1,3}$, Xiaoguang Wang$^{1,2}$, Franco
Nori$^{1,4}$}

\affiliation{$^{1}$Advanced Science Institute, RIKEN, Wako-shi, Saitama 351-0198,
Japan\\
 $^{2}$Zhejiang Institute of Modern Physics, Department of Physics,
Zhejiang University, Hangzhou 310027, China\\
 $^{3}$Department of Physics, Hangzhou Normal University, Hangzhou
310027, China\\
 $^{4}$Physics Department, The University of Michigan, Ann Arbor,
MI 48109-1040, USA }
\begin{abstract}
We derive a set of hierarchical equations for qubits interacting with
a Lorentz-broadened cavity mode at zero temperature, without using
the rotating-wave, Born, and Markovian approximations. We use this
exact method to reexamine the entanglement dynamics of two qubits
interacting with a common bath, which was previously solved only under
the rotating-wave and single-excitation approximations. With the exact
hierarchy equation method used here, double excitations due to counter-rotating-wave
terms are found to have remarkable effects on the dynamics and the
steady state entanglement. 
\end{abstract}

\pacs{03.65.Ta, 03.65.Ud, 03.65.Yz}

\maketitle

\section{Introduction}

Decoherence is one of the most important problem in quantum information
processing \cite{Breuer_book}. The description of this difficult
problem usually involves various approximations. During the dynamic
evolution, the system and the bath are mixed, and a perturbative treatment
is required such that we can trace out the degrees of freedom of the
bath. This perturbation is known as the Born approximation. Moreover,
if the time scale of the bath is much shorter than that of the system,
the Markovian approximation is often applied.

An effective method that avoids the above two approximations was developed
by Tanimura \textit{et al}. \cite{Tanimura1989.1991,Ishizaki.2005.2007,Tanimura2006},
who established a set of hierarchical equations \cite{Tanimura2006}
that includes all orders of system-bath interactions. The derivation
of the hierarchy equations requires that the time-correlation function
of the bath can be decomposed into a set of exponential functions
\cite{Tanimura2006}. At finite temperature, this requirement is fulfilled
if the system-bath coupling can be described by a Drude spectrum.
The hierarchy equation method is successfully used in describing quantum
dynamics of chemical and biophysical systems \cite{Ishizaki.2005.2007,Tanimura2006,YJ_Yan,Ishizaki_FMO},
such as the light-harvesting complexes \cite{Ishizaki_FMO}, of which
the temperature of the environment is high enough, and the coupling
between the system and the environment is too strong to enable a Born
approximation. However, the powerful hierarchy equation method was
seldom used in studying decoherence effects in quantum information
\cite{JQ_You}. Firstly, the operating temperature of qubit devices
is very low. If we use the Drude spectrum, a numerical difficulty
arises since the time-correlation function of the bath should be decomposed
into a very large set of exponential functions \cite{Ishizaki.2005.2007}.
Actually, the temperature of qubit devices is low enough that we can
use a zero-temperature environment to model the decoherence. Secondly,
the Drude spectrum is not quite general in qubit devices, especially
when the qubit is placed in a cavity, and its environment is usually
modeled by a Lorentz-broadened cavity mode.

In this paper we find that the hierarchy equation can also be derived
at zero-temperature if we employ a Lorentz-type system-bath coupling
spectrum. The set of hierarchy equations derived here provides an
exact treatment of decoherence, and employs neither the rotating-wave,
Born, nor Markovian approximations. System-bath correlations are here
fully accounted during the entire time evolution, as compared to traditional
master equation treatments, the correlations are truncated to second
order. High-order correlations are shown \cite{Tanimura2010PRL} to
be very important, even producing a totally different physics. Moreover,
the hierarchy equation we derive here is found to be effective in
the single-mode case, and is a promising method for studying strong-
and ultrastrong- coupling physics \cite{JQ_You,StrongCoupling_QED}.

We use the hierarchy equation method to study a model of two qubits
interacting with a common bosonic bath, which is widely considered
in studying decoherence-free subspace \cite{DFS} and bipartite entanglement
dynamics \cite{t_yu}. This model was solved exactly \cite{Breuer_book,Maniscalco2008PRL}
under the rotating-wave approximation (RWA). It is
not surprising that entanglement can be generated for a separable
initial state, since the bath induces an effective qubit-qubit interaction.
Another observation based on the RWA 
lies in the steady-state entanglement,
which is determined only by the overlap between the initial state
and the decoherence-free state, independent of the system-bath coupling
\cite{Maniscalco2008PRL}. This is because the dynamics of the qubit
is restricted to a single-excitation subspace. However, when the counter-rotating
terms are accounted, double excitation occurs and the steady-state
entanglement vanishes for certain system-bath couplings. We will demonstrate
this observation below. ~ ~ ~

%:-------Hierarchy Eqn--------

\section{ Hierarchy equation method}

Here we first consider qubits interacting with a bosonic bath, also
known as the spin-boson model: %: eq:Ham
\begin{equation}
H=H_{S}+H_{B}+H_{\text{Int}},\label{eq:Ham}
\end{equation}
 where $H_{S}$ is the free Hamiltonian of the qubit and (with $\hbar=1$)
%: eq:Ham_B_Int
\begin{eqnarray}
H_{B} & = & \sum_{k}\omega_{k}b_{k}^{\dagger}b_{k},\nonumber \\
H_{\text{Int}} & = & \sum_{k}V\left(g_{k}b_{k}+g_{k}^{*}b_{k}^{\dagger}\right),\label{eq:Ham_B_Int}
\end{eqnarray}
 where $b_{k}^{\dagger}$ and $b_{k}$ are the creation and annihilation
operators of the bath, $V$ is the operator of the qubit, while $g_{k}$
is the coupling strength between the qubit and the $k$th mode of
the bath.

The exact dynamics of the system in the interaction picture can be
derived as \cite{Tanimura2006} %:	widetext
\begin{widetext} %: eq:rhot_general
 
\begin{eqnarray}
\rho_{S}^{(I)}\left(t\right) & = & \mathcal{T}\exp\left\{ -\int_{0}^{t}\!\!\! dt_{2}\!\int_{0}^{t_{2}}\!\!\!\!\! dt_{1}V\left(t_{2}\right)^{\times}\!\left[C^{R}\left(t_{2}-t_{1}\right)V\left(t_{1}\right)^{\times}\!\!+iC^{I}\left(t_{2}-t_{1}\right)V\left(t_{1}\right)^{\circ}\right]\right\} \rho_{S}\left(0\right),\label{eq:rhot_general}
\end{eqnarray}
 \end{widetext} if the qubit and bath are initially in a separable
state, i.e. $\rho\left(0\right)=\rho_{S}\left(0\right)\otimes\rho_{B}$,
where $\rho_{B}=\exp\left(-\beta H_{B}\right)/Z_{B}$ is the initial
state of the bath, with $\beta=1/T$ (with $k_{B}=1$) and $Z_{B}$
is the partition function. In Eq.~(\ref{eq:rhot_general}), $\mathcal{T}$
is the chronological time-ordering operator, which orders the operators
inside the integral such that the time arguments increase from right
to left. Two superoperators are introduced, $A^{\times}B\equiv\left[A,\, B\right]=AB-BA$
and $A^{\circ}B\equiv\left\{ A,\, B\right\} =AB+BA$. Also, $C^{R}\left(t_{2}-t_{1}\right)$
and $C^{I}\left(t_{2}-t_{1}\right)$ are the real and imaginary parts
of the bath time-correlation function 
\begin{equation}
C\left(t_{2}-t_{1}\right)\equiv\left\langle B\left(t_{2}\right)B\left(t_{1}\right)\right\rangle =\text{Tr}\left[B\left(t_{2}\right)B\left(t_{1}\right)\rho_{B}\right],
\end{equation}
 respectively, and %: eq:bath_corr
\begin{equation}
B\left(t\right)=\sum_{k}\left(g_{k}b_{k}e^{-i\omega_{k}t}+g_{k}^{*}b_{k}^{\dagger}e^{i\omega_{k}t}\right).\label{eq:bath_corr}
\end{equation}
 Equation (\ref{eq:rhot_general}) is difficult to solve directly,
due the time-ordered integral. An effective method for this problem
was developed \cite{Tanimura1989.1991,Ishizaki.2005.2007,Tanimura2006,YJ_Yan}
by solving a set of hierarchy equations, such as the form of Eq.~(\ref{eq:hier_eq}).
The hierarchy equations are obtained by repeatedly taking the derivative
of the right-hand side when the system-bath coupling is described
by the Drude spectrum $J\left(\omega\right)=\frac{\omega}{\pi}\frac{2\eta\omega_{c}}{\omega^{2}+\omega_{c}^{2}}$
at finite temperatures, where $\eta$ is the reorganization energy,
and $\omega_{c}$ is the decay rate of the bath correlation function.
A key condition in deriving the hierarchy equation is that, with the
Drude spectrum, the correlation function (\ref{eq:bath_corr}) can
be decomposed into a sum of exponential functions of time as $C\left(t_{2}-t_{1}\right)=\sum_{k}f_{k}\exp\left(-\gamma_{k}\right)$,
where $\gamma_{k}=\frac{2\pi k}{\beta}\left(1-\delta_{k,0}\right)+\omega_{c}\delta_{k,0}$
are the Matsubara frequencies. The hierarchy equation method enables
a rigorous study of decoherence-related effects in chemical physics
and biophysics \cite{Ishizaki_FMO}. In such systems, the coupling
strength between the system and bath is not always weak, and the temperature
$T$ is so high that only a few Matsubara terms could provide enough
numerical precision \cite{Ishizaki.2005.2007}. However, the number
of Matsubara terms in the expansion increases with decreasing temperature,
which is difficult to handle numerically. This problem becomes serious
when we consider qubit devices, which are generally prepared in nearly
zero-temperature environments, and thus prevent the use of the original
hierarchy equation method. Fortunately, the exponential decay of bath
correlation functions at zero temperature occurs in many quantum computing
devices, such as cavity-qubit systems, where the coupling spectrum
between the qubits and cavity modes is usually Lorentz type, but not
Drude type, so in that case the hierarchy method can also be applied.

Now we consider qubits interacting with a single mode of the cavity,
with frequency $\omega_{0}$. Due to the imperfection of the cavity,
the single mode is broadened and the qubit-cavity coupling spectrum
becomes Lorentz-type 
\begin{equation}
J\left(\omega\right)=\frac{1}{\pi}\frac{\lambda\gamma}{\left(\omega-\omega_{0}\right)^{2}+\gamma^{2}},
\end{equation}
 where $\lambda$ reflects the system-bath coupling strength, $\gamma$
is the broadening width of the cavity mode, and $\tau_{c}=1/\gamma$
is the lifetime of the mode. At $T=0$, if the cavity is initially
in a vacuum state $\otimes_{k}|0_{k}\rangle$, the time-correlation
function (\ref{eq:bath_corr}) becomes %: eq:Lor_corr
\begin{equation}
C\left(t_{2}-t_{1}\right)=\lambda\exp\left[-\left(\gamma+i\omega_{0}\right)|t_{2}-t_{1}|\right],\label{eq:Lor_corr}
\end{equation}
 which is an exponential form that we need to use for the hierarchy
equations. In the single-mode limit, $\gamma=0$ and $C\left(t_{2}-t_{1}\right)=\lambda\exp\left(-i\omega_{0}\left|t_{2}-t_{1}\right|\right)$,
and we see that $\lambda$ is related to the square of the Rabi oscillation
frequency.

To derive the hierarchy equation in a convenient form, we further
write the real and imaginary parts of the time-correlation function
(\ref{eq:Lor_corr}) as 
\begin{equation}
C^{R}\left(t\right)=\sum_{k=1}^{2}\frac{\lambda}{2}e^{-\nu_{k}t},\; C^{I}\left(t\right)=\sum_{k=1}^{2}\left(-1\right)^{k}\frac{\lambda}{2i}e^{-\nu_{k}t},
\end{equation}
 where $\nu_{k}=\gamma+(-1)^{k}i\omega_{0}$. Then, following procedures
shown in Appendix A and Ref.~\cite{Tanimura1989.1991,Tanimura2006}, the hierarchy equations
of the qubits are obtained 
\begin{eqnarray}
\frac{\partial}{\partial t}\varrho_{\vec{n}}\left(t\right) & = & -\left(iH_{S}^{\times}+\vec{n}\cdot\vec{\nu}\right)\varrho_{\vec{n}}\left(t\right)-i\sum_{k=1}^{2}V^{\times}\varrho_{\vec{n}+\vec{e}_{k}}\left(t\right)\nonumber \\
 & - & i\frac{\lambda}{2}\sum_{k=1}^{2}n_{k}\left[V^{\times}+\left(-1\right)^{k}V^{\circ}\right]\varrho_{\vec{n}-\vec{e}_{k}}\left(t\right),\label{eq:hier_eq}
\end{eqnarray}
 where the subscript $\vec{n}=\left(n_{1},\, n_{2}\right)$ is a two-dimensional
index, with $n_{1\left(2\right)}\ge0$, and $\rho_{S}\left(t\right)\equiv\varrho_{\left(0,0\right)}\left(t\right)$.
The vectors $\vec{e}_{1}=\left(1,\,0\right)$, $\vec{e}_{2}=\left(0,\,1\right)$,
and $\vec{\nu}=\left(\nu_{1},\,\nu_{2}\right)=\left(\gamma-i\omega_{0},\,\gamma+i\omega_{0}\right)$.
We emphasize that $\varrho_{\vec{n}}\left(t\right)$ with $\vec{n}\ne\left(0,\,0\right)$
are auxiliary operators introduced only for the sake of computing,
they are not density matrices, and are all set to be zero at $t=0$.
The hierarchy equations are a set of linear differential equations,
and can be solved by using the Runge-Kutta method. 
The contributions of the bath to the dynamics of the system, including both dissipation and Lamb shift, are
fully contained in the hierarchy equation (\ref{eq:hier_eq}).
The Lamb shift term \cite{Hoeppe2012}, which is related to the imaginary part of 
the bath correlation function, can be written explicitly in the common
non-Markovian equations. 
Since the real and imaginary parts of 
the bath correlation function are taken into considered here,
the effects of the Lamb shift exist in the hierarchy equations,
although not in an explicit form.

For numerical computations,
the hierarchy equation (\ref{eq:hier_eq}) must be truncated for large enough $\vec{n}$.
We can increase the hierarchy order $\vec{n}$ until the results of $\rho_S(t)$
converge.
The terminator of the hierarchy equation is 
\begin{eqnarray}
\frac{\partial}{\partial t}\varrho_{\vec{N}}\left(t\right) & = &
-\left(iH_{S}^{\times}+\vec{N}\cdot\vec{\nu}\right)\varrho_{\vec{n}}\left(t\right)\nonumber \\
 & - &
 i\frac{\lambda}{2}\sum_{k=1}^{2}n_{k}\left[V^{\times}+\left(-1\right)^{k}V^{\circ}\right]\varrho_{\vec{N}-\vec{e}_{k}}\left(t\right),\nonumber
 \\
 \label{eq:hier_trunc}
\end{eqnarray}
where we dropped the deeper auxiliary operators $\varrho_{\vec{N}+\vec{e}_k}$.
The numerical results in this paper were all tested and converged, and the
density matrix $\rho_S(t)$ is positive.

%:-------Common bath--------
%: fig1
\begin{figure}[ptb]
\begin{centering}
\includegraphics[width=8cm]{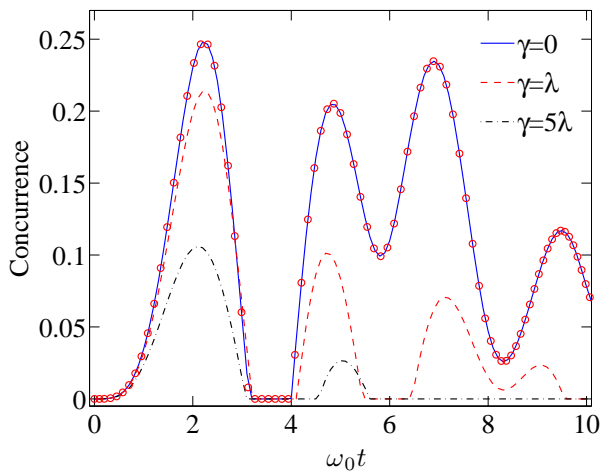} \caption{(Color online) Concurrence versus time for the initial state $|\psi(0)\rangle=|0\rangle_{1}|0\rangle_{2}$
with different values of $\gamma$. Here $\lambda=0.1\omega_{0}$
is in the strong-coupling regime. In the single-mode limit, $\gamma=0$,
the result of the hierarchy equation (solid) coincides with direct
numerical calculations (circles). The entanglement suddenly vanishes
and revivals are observed. When increasing $\gamma$, the oscillations
and the maximum entanglement are suppressed. Under the RWA, the initial state does not evolve, and the entanglement
stays at zero.}
\label{fig1}
\par\end{centering}
\end{figure}

\section{ Entanglement of two qubits in a common bath}

Below we apply the hierarchy equation (\ref{eq:hier_eq}) to a widely
studied model: two qubits interacting with a common bath. The model
is used to study decoherence-free space \cite{DFS}, bath induced
entanglement \cite{Maniscalco2008PRL}, and other related topics \cite{t_yu}.
In previous works \cite{Maniscalco2008PRL,Francia2009PRA,Mazzola2009PRA},
the RWA was used, and the exact dynamics could
only be found in a single-excitation subspace. Without using the RWA, the model was also studied \cite{XF_Cao,JingJun,Benatti}
in a perturbative way. However, if the system-bath coupling becomes
strong enough, which is explored in recent experiments \cite{StrongCoupling_QED},
both the RWA and perturbation methods fail.
Therefore the hierarchy method is very suitable in such conditions.

Consider two qubits interacting with a common bosonic bath. The system
Hamiltonian in Eq.~(\ref{eq:Ham}) is now given by 
\begin{equation}
H_{S}=\frac{\omega_{1a}}{2}\sigma_{1z}+\frac{\omega_{2a}}{2}\sigma_{2z},
\end{equation}
and below we consider $\omega_{1a}=\omega_{2a}=\omega_{0}$, i.e.,
the resonant case. The system operator in Eq.~(\ref{eq:Ham_B_Int})
is set by $V=\alpha_{1}\sigma_{1x}+\alpha_{2}\sigma_{2x}$, and for
simplicity, we consider $\alpha_{1}=\alpha_{2}=1$. This model is
exactly solvable \cite{Maniscalco2008PRL} when the RWA is applied and the initial state is of the form 
\begin{equation}
|\psi\left(0\right)\rangle=\left[c_{1}\left(0\right)|1\rangle_{1}|0\rangle_{2}+c_{2}\left(0\right)|0\rangle_{1}|1\rangle_{2}\right]\bigotimes_{k}|0_{k}\rangle.
\end{equation}
 The time evolution is then given by 
\begin{eqnarray}
|\psi\left(t\right)\rangle & = & \left[c_{1}\left(t\right)|1\rangle_{1}|0\rangle_{2}+c_{2}\left(t\right)|0\rangle_{1}|1\rangle_{2}\right]\bigotimes_{k}|0_{k}\rangle\nonumber \\
 &  & +\sum_{k}c_{k}\left(t\right)|0\rangle_{1}|0\rangle_{2}|1_{k}\rangle,
\end{eqnarray}
 where $|1_{k}\rangle$ denotes that only the $k$th mode of the bath
is excited. The explicit forms of $c_{1}\left(t\right)$ and $c_{2}\left(t\right)$
are given in Ref.~\cite{Maniscalco2008PRL}. The time evolution of
the density matrix is 
\begin{equation}
\rho\left(t\right)\!=\!\!\left(\begin{array}{cccc}
0 & \!0 & \!0 & \!\!\!0\\
0 & \!\left|c_{1}\!\left(t\right)\right|^{2} & \! c_{1}\!\left(t\right)\! c_{2}^{*}\!\left(t\right) & \!\!\!0\\
0 & \! c_{2}\!\left(t\right)\! c_{1}^{*}\!\left(t\right) & \!\left|c_{2}\!\left(t\right)\right|^{2} & \!\!\!0\\
0 & 0 & 0 & \!\!\!1\!-\!\left|c_{1}\!\left(t\right)\right|^{2}\!-\!\left|c_{2}\!\left(t\right)\right|^{2}
\end{array}\right),
\end{equation}
 which is obviously restricted to a single-excitation space, and thus
the concurrence of the above density matrix is 
\begin{equation}
C\left(t\right)=2\left|c_{1}\left(t\right)c_{2}^{*}\left(t\right)\right|.\label{eq:conc1}
\end{equation}

We first compare the above results with our hierarchy method for
the initial state $|\psi(0)\rangle=|0\rangle_{1}|0\rangle_{2}$ ,
shown in Fig.~\ref{fig1}. The system-bath coupling is set by $\lambda=0.1\omega_{0}$,
which already enters the strong-coupling regime. Such an initial state
does not evolve under the RWA, and then no
entanglement will be produced. However, from Fig.~\ref{fig1} we
observe the generation of considerable entanglement, even with large
$\gamma$. The RWA fails in predicting the
real physics. Since the coupling is strong, the oscillation for small
$\gamma$ case is drastic. The sudden vanishing and revival of entanglement
were observed, and with increasing of $\gamma$, the oscillations
of the concurrence were suppressed. It should be emphasized that,
when $\gamma=0$, the results obtained by the hierarchy equations
coincide with our exact numerical results obtained by solving the
single-mode Hamiltonian directly. Therefore, by using a unified method,
we can study the dynamics of the system interacting with a bath from
the single-mode to multimode regime. 

%: fig2
\begin{figure}[tb]
\begin{centering}
\includegraphics[width=9cm]{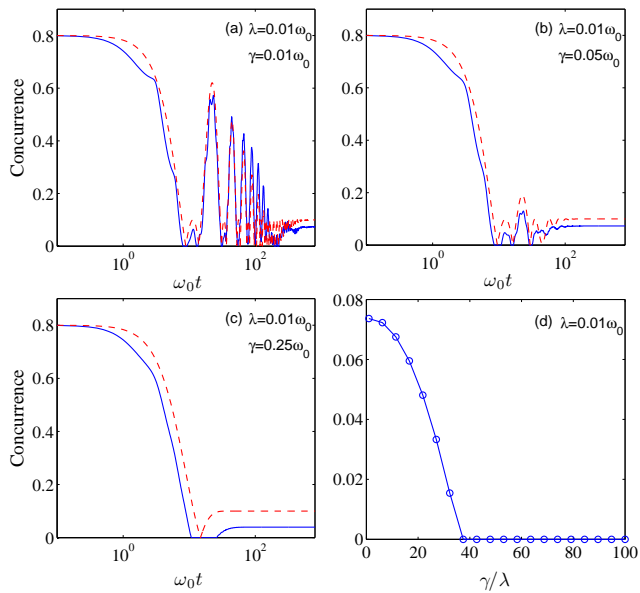} \caption{(Color online) (a)-(c): Dynamics of the concurrence computed with
the hierarchy equations (solid) and exact results under the RWA (dashed). The coupling strength $\lambda=0.01\omega_{0}$
is so strong that the RWA is invalid. With
increasing $\gamma$, the non-Markovian oscillations disappear. It
is interesting that the steady-state entanglement obviously depends
on $\gamma$, which cannot be predicted by using the RWA. The steady-state entanglement is shown in (d), and
it vanishes if $\gamma$ is larger than a critical value. }
\label{fig2} 
\par\end{centering}
\end{figure}

Another interesting result here is about the steady-state entanglement.
Under the RWA, the dynamics is in the single-excitation
subspace, only two states are independent, $|\varphi_{\pm}\rangle=\left(|0\rangle_{1}|1\rangle_{2}\pm|1\rangle_{1}|0\rangle\right)/\sqrt{2}$.
The state $|\varphi_{-}\rangle$ is decoherence-free; this means that
if the initial state has non-vanishing overlap with $|\varphi_{-}\rangle$,
the steady state is entangled, and the concurrence becomes 
\begin{equation}
C\left(t\rightarrow\infty\right)=C\left(|\varphi_{-}\rangle\right)\left|\left\langle \varphi_{-}|\psi\left(0\right)\right\rangle \right|^{2}=\left|\left\langle \varphi_{-}|\psi\left(0\right)\right\rangle \right|^{2},\label{eq:conc_steady}
\end{equation}
 which is independent of the system-bath coupling strength $\lambda$
and the bath-decay rate $\gamma$. However, if $\lambda$ is not very
small, although $|\varphi_{-}\rangle$ is also decoherence-free, Eq.~(\ref{eq:conc_steady})
should be reexamined by using a more rigorous treatment, since double
excitations need to be accounted. 
Actually, the reliability of the RWA was discussed in many 
literatures \cite{Seke1993,Intravaia2003,Nesi2007,Werlang2008,Wang2008,XF_Cao,JingJun,Benatti,Altintas2012}.
As shown in Refs.~\cite{Seke1993,Altintas2012}, counter-rotating-wave terms can
induced a significant shift in the population of the steady state even
in the bad-cavity case. 

In Fig.~\ref{fig2}, we show the
results given by the hierarchy method. The initial state there is
$|\psi(0)\rangle=(2|0\rangle_{1}|1\rangle_{2}+|1\rangle_{1}|0\rangle_{2})/\sqrt{5}$.
According to Eq.~(\ref{eq:conc_steady}), the concurrence of the
steady state is $0.1$. We can see in Fig.~\ref{fig2}(a)-(c) that
increasing $\gamma$ the concurrence of the steady state decreases.
In Fig.~\ref{fig2}(d), we show that for a given $\lambda=0.01\omega_{0}$,
the steady-state entanglement vanishes when $\gamma$ is larger than
a critical value. This reflects the importance of the counter-rotating-wave
terms, which break the single-excitation condition and give a totally
different steady-state entanglement. 
Similar results are obtained in Ref.~\cite{Altintas2012}, where
the increase of the cavity decay rate is found to decrease the maximum of
induced entanglement, and the steady state that computed without RWA has no 
entanglement but finite discord. This simple example indicates
that some exact results previously obtained under the RWA need to be reexamined.
%
%:-------Conclusion--------
~

\section{Conclusion}

In summary, we derive a set of hierarchy equations at zero temperature
with a Lorentz spectrum. This set of equations is very suitable for
qubit-cavity systems, especially when the interaction is so strong
that the RWA and perturbative methods break
down. It even works well when the bath has only one single mode. Moreover,
this equation is very flexible. For example, if the qubits interact
with several cavity modes, each broadened into a Lorentz form, then
the bath correlation functions can also be expanded as several exponential
functions. Thus the form of the hierarchy equations remains. The hierarchy
equations are applied to reexamine the dynamics of two qubits interacting
with a common bath. Previous works usually employed the RWA, 
and the results were restricted to the single-excitation
space. This is not the case in this paper, since we do not use the
RWA, and the counter-rotating-wave terms will
cause double-excitations. We found that the steady-state entanglement
depends on the system-bath coupling spectrum. For a given coupling
strength $\lambda$, there will be no steady state entanglement when
$\gamma$ is larger than a critical value. The exact dynamics exhibits
a totally different physics, compared to the RWA
model, which motivates the re-examination of many previous approximate
studies.

~

\section{Acknowledgements}

X. Wang acknowledges support from the NFRPC with Grant No.~2012CB921602
and NSFC with grant No.~11025527 and 10935010. FN acknowledges partial
support from the LPS, NSA, ARO, NSF grant No.~0726909, JSPS-RFBR
contract No.~09-02-92114, Grant-in-Aid for Scientific Research (S),
MEXT Kakenhi on Quantum Cybernetics, and the JSPS through its FIRST
program. Jian Ma acknowledges support from the Scholarship Grant for
Excellent Doctoral Student granted by the Ministry of Education. Zhe
Sun acknowledges support from the National Nature Science Foundation
of China with Grant No.~11005027; the National Science Foundation
of Zhejiang Province with Grant No.~Y6090058, and the Program for
Excellent Young Teachers in Hangzhou Normal University with Grant
No.~HNUEYT 2011-01-01.

\appendix

\section{Derivation of the hierarchy equations}

Below, we derive the hierarchy equations. Firstly, inserting the correlation
function (\ref{eq:Lor_corr}) into Eq.~(\ref{eq:rhot_general}),
we find
\begin{widetext}
\begin{eqnarray}
\rho_{S}\left(t\right) & = & U\left(t\right)\left\{ \mathcal{T}\exp\left(-\int_{0}^{t}\!\!\! dt_{2}\!\!\int_{0}^{t_{2}}\!\!\!\! dt_{1}V\left(t_{2}\right)^{\times}\sum_{k=1}^{2}\frac{\lambda}{2}e^{-\nu_{k}\left(t_{2}-t_{1}\right)}\left[V\left(t_{1}\right)^{\times}\!\!+\left(-1\right)^{k}V\left(t_{1}\right)^{\circ}\right]\right)\rho_{S}\left(0\right)\right\} U^{\dagger}\left(t\right)\nonumber \\
 & = & U\left(t\right)\left\{ \mathcal{T}\exp\left[\int_{0}^{t}\!\!\! dt_{2}\!\!\int_{0}^{t_{2}}\!\!\!\! dt_{1}\Phi\left(t_{2}\right)\sum_{k=1}^{2}e^{-\nu_{k}\left(t_{2}-t_{1}\right)}\Theta_{k}\left(t_{1}\right)\right]\rho_{S}\left(0\right)\right\} U^{\dagger}\left(t\right),\label{eq:rhot2}
\end{eqnarray}
 where $U(t)=\exp[-i\,(H_S+H_B)t]$, and the two new superoperators 
\begin{eqnarray}
\Phi\left(t\right) & = & -iV\left(t\right)^{\times},\nonumber \\
\Theta_{k}\left(t\right) & = & -\frac{i\lambda}{2}\left[V\left(t\right)^{\times}+\left(-1\right)^{k}V\left(t\right)^{\circ}\right]\label{eq:PhiTheta}
\end{eqnarray}
 in order to make the following discussion clearer and simpler. Equation
(\ref{eq:rhot2}) is a time-ordered integral equation, which is not
easy to solve directly. The idea of the hierarchy equation method
\cite{Tanimura1989.1991,Tanimura2006} is to transform such an integral
equation to a group of ordinary differential equations. The derivation
of the hierarchy equations is straightforward: taking the time derivative
of Eq.~(\ref{eq:rhot2}) repeatedly.

We first take the time derivative of Eq.~(\ref{eq:rhot2}) and obtain
\begin{equation}
\frac{\partial}{\partial t}\rho_{S}\left(t\right)=-iH_{S}^{\times}\rho_{S}\left(t\right)+\Phi\sum_{k=1}^{2}\mathcal{F}_{k}\left(t\right),\label{eq:drho_sys1}
\end{equation}
 where 
\begin{equation}
\mathcal{F}_{k}\left(t\right)=U\left(t\right)\mathcal{T}\left\{ \int_{0}^{t}d\tau e^{-\nu_{k}\left(t-\tau\right)}\Theta_{k}\left(\tau\right)\exp\left[\int_{0}^{t}\, dt_{2}\int_{0}^{t_{2}}\, dt_{1}\Phi\left(t_{2}\right)\sum_{n=1}^{2}e^{\nu_{n}\left(t_{2}-t_{1}\right)}\Theta_{n}\left(t_{1}\right)\right]\right\} \rho_{S}\left(0\right)U\left(t\right)^{\dagger}.\label{eq:Fk}
\end{equation}
Thus the solution of $\rho_{S}\left(t\right)$ is determined by (i)
its own free evolution, (ii) the dynamics of $\mathcal{F}_{k}\left(t\right)$.
The initial condition of $\mathcal{F}_{k}\left(t\right)$ is 
\begin{equation}
\mathcal{F}_{k}\left(0\right)=0,
\end{equation}
 which is a direct result of Eq.~(\ref{eq:Fk}). To solve for $\mathcal{F}_{k}\left(t\right)$,
we need its differential equation. Before taking the time derivative
of $\mathcal{F}_{k}\left(t\right)$, we first introduce the following
useful notations \cite{Tanimura2006}: 
\begin{eqnarray}
\varrho_{\left(0,0\right)}\left(t\right) & \equiv & \rho_{S}\left(t\right),\nonumber \\
\varrho_{\left(1,0\right)}\left(t\right) & \equiv & \mathcal{F}_{1}\left(t\right),\nonumber \\
\varrho_{\left(0,1\right)}\left(t\right) & \equiv & \mathcal{F}_{2}\left(t\right).
\end{eqnarray}
 Then Eq.~(\ref{eq:drho_sys1}) can be rewritten as 
\begin{eqnarray}
\frac{\partial}{\partial t}\rho_{S}\left(t\right) & = & -iH_{S}^{\times}\rho_{S}\left(t\right)+\Phi\left[\varrho_{\left(1,0\right)}\left(t\right)+\varrho_{\left(0,1\right)}\left(t\right)\right],\nonumber \\
 & = & -iH_{S}^{\times}\rho_{S}\left(t\right)+\Phi\sum_{k=1}^{2}\varrho_{\left(0,0\right)+\vec{e}_{k}}\left(t\right),\label{eq:drho_sys}
\end{eqnarray}
 where $\vec{e}_{1}=\left(1,\,0\right)$and $\vec{e}_{2}=\left(0,\,1\right)$.

The differential equations of $\varrho_{\left(1,0\right)}\left(t\right)$
and $\varrho_{\left(0,1\right)}\left(t\right)$ are obtained as 
\begin{eqnarray}
\frac{\partial}{\partial t}\varrho_{\left(1,0\right)}\left(t\right) & = & -\left(iH_{S}^{\times}+\nu_{1}\right)\varrho_{\left(1,0\right)}\left(t\right) +\Phi\sum_{k=1}^{2}\varrho_{\left(1,0\right)+\vec{e}_{k}}\left(t\right)+\Theta_{1}\varrho_{\left(0,0\right)}\left(t\right)\nonumber \\
\frac{\partial}{\partial t}\varrho_{\left(0,1\right)}\left(t\right) & = & -\left(iH_{S}^{\times}+\nu_{2}\right)\varrho_{\left(0,1\right)}\left(t\right)
+\Phi\sum_{k=1}^{2}\varrho_{\left(0,1\right)+\vec{e}_{k}}\left(t\right)+\Theta_{2}\varrho_{\left(0,0\right)}\left(t\right),\nonumber \\
~
\end{eqnarray}
 where we find three new auxiliary matrices, 
\begin{eqnarray}
\varrho_{\left(2,0\right)}\left(t\right) & = & U\left(t\right)\mathcal{T}\bigg\{\left[\int_{0}^{t}d\tau e^{-\nu_{1}\left(t-\tau\right)}\Theta_{1}\left(\tau\right)\right]^{2}\nonumber \\
 &  & \times\exp\left[\int_{0}^{t}\, dt_{2}\int_{0}^{t_{2}}\, dt_{1}\Phi\left(t_{2}\right)\sum_{k=1}^{2}e^{\nu_{k}\left(t_{2}-t_{1}\right)}\Theta_{k}\left(t_{1}\right)\right]\bigg\}\rho_{S}\left(0\right)U\left(t\right)^{\dagger},\\
\varrho_{\left(1,1\right)}\left(t\right) & = & U\left(t\right)\mathcal{T}\bigg\{\left[\int_{0}^{t}d\tau e^{-\nu_{1}\left(t-\tau\right)}\Theta_{1}\left(\tau\right)\right]\left[\int_{0}^{t}d\tau e^{-\nu_{2}\left(t-\tau\right)}\Theta_{2}\left(\tau\right)\right]\nonumber \\
 &  & \times\exp\left[\int_{0}^{t}\, dt_{2}\int_{0}^{t_{2}}\, dt_{1}\Phi\left(t_{2}\right)\sum_{k=1}^{2}e^{\nu_{k}\left(t_{2}-t_{1}\right)}\Theta_{k}\left(t_{1}\right)\right]\bigg\}\rho_{S}\left(0\right)U\left(t\right)^{\dagger},\\
\varrho_{\left(0,2\right)}\left(t\right) & = & U\left(t\right)\mathcal{T}\bigg\{\left[\int_{0}^{t}d\tau e^{-\nu_{2}\left(t-\tau\right)}\Theta_{2}\left(\tau\right)\right]^{2}\nonumber \\
 &  & \times\exp\left[\int_{0}^{t}\, dt_{2}\int_{0}^{t_{2}}\, dt_{1}\Phi\left(t_{2}\right)\sum_{k=1}^{2}e^{\nu_{k}\left(t_{2}-t_{1}\right)}\Theta_{k}\left(t_{1}\right)\right]\bigg\}\rho_{S}\left(0\right)U\left(t\right)^{\dagger}.
\end{eqnarray}
 By repeating the above procedures, we find 
\begin{equation}
\frac{\partial}{\partial t}\varrho_{\vec{n}}\left(t\right)=-\left(iH_{S}^{\times}+\vec{n}\cdot\vec{\nu}\right)\varrho_{\vec{n}}\left(t\right)+\Phi\sum_{k=1}^{2}\varrho_{\vec{n}+\vec{e}_{k}}\left(t\right)+\sum_{k=1}^{2}n_{k}\Theta_{k}\varrho_{\vec{n}-\vec{e}_{k}}\left(t\right),\label{eq:hier1}
\end{equation}
 where $\vec{n}=\left(n_{1},\, n_{2}\right)$ is a two-dimensional
index, with $n_{1(2)}\ge0$. The two-dimensional vector $\vec{\nu}=\left(\nu_{1},\,\nu_{2}\right)=\left(\gamma-i\omega_{0},\,\gamma+i\omega_{0}\right)$.
The auxiliary matrix is 
\begin{eqnarray}
\varrho_{\vec{n}}\left(t\right) & = & U\left(t\right)\mathcal{T}\bigg\{\left[\int_{0}^{t}d\tau e^{-\nu_{1}\left(t-\tau\right)}\Theta_{1}\left(\tau\right)\right]^{n_{1}}\left[\int_{0}^{t}d\tau e^{-\nu_{2}\left(t-\tau\right)}\Theta_{2}\left(\tau\right)\right]^{n_{2}}\nonumber \\
 &  & \times\exp\left[\int_{0}^{t}\, dt_{2}\int_{0}^{t_{2}}\, dt_{1}\Phi\left(t_{2}\right)\sum_{k=1}^{2}e^{\nu_{k}\left(t_{2}-t_{1}\right)}\Theta_{k}\left(t_{1}\right)\right]\bigg\}\rho_{S}\left(0\right)U\left(t\right)^{\dagger}.
\end{eqnarray}
 Inserting Eq.~(\ref{eq:PhiTheta}) into Eq.~(\ref{eq:hier1}),
we obtain the explicit form of the hierarchy equation as 
\begin{eqnarray}
\frac{\partial}{\partial t}\varrho_{\vec{n}}\left(t\right) & = & -\left(iH_{S}^{\times}+\vec{n}\cdot\vec{\nu}\right)\varrho_{\vec{n}}\left(t\right)\nonumber \\
 &  & -i\sum_{k=1}^{2}V^{\times}\varrho_{\vec{n}+\vec{e}_{k}}\left(t\right)\nonumber \\
 &  & -i\frac{\lambda}{2}\sum_{k=1}^{2}n_{k}\left[V^{\times}+\left(-1\right)^{k}V^{\circ}\right]\varrho_{\vec{n}-\vec{e}_{k}}\left(t\right).\nonumber \\
~
\end{eqnarray}
 The initial conditions are 
\[
\varrho_{\vec{n}}\left(0\right)=\begin{cases}
\rho_{S}\left(0\right), & \mbox{for }n_{1}=n_{2}=0,\\
0, & \mbox{for }n_{1}>0,\ n_{2}>0.
\end{cases}
\]
 Although the explicit form of $\varrho_{\vec{n}}\left(t\right)$
is complicated, we need only to focus on its differential equations,
which can be solved directly by using the traditional Runge-Kutta
method. 
\end{widetext}
%:-------References--------

%:Breuer_book


\begin{thebibliography}{References}
\bibitem{Breuer_book} H.-P.~Breuer and F.~Petruccione, \textit{The
Theory of Open Quantum Systems} (Oxford Univ. Press, New York, 2002).
%:Tanimura1989.1991 (First hierarchy)
 
%:Tanimura1989.1991 (First hierarchy)
\bibitem{Tanimura1989.1991}
Y.~Tanimura and R.~Kubo, J.~Phys.~Soc.~Jpn. {\bf 58}, 101 (1989);
Y.~Tanimura, Phys.~Rev.~A {\bf 41}, 6676 (1990); Y.~Tanimura and P.~G.~Wolynes, Phys.~Rev.~A {\bf 43}, 4131 (1991);
Y.~Tanimura and S.~Mukamel, J.~Phys.~Soc.~Jpn. {\bf 63}, 66 (1993);
M.~Tanaka and Y.~Tanimura, J.~Phys.~Soc.~Jpn. {\bf 78}, 073802 (2009);
M.~Tanaka and Y.~Tanimura, J.~Chem.~Phys. {\bf 132} 214502 (2010).

%:Ishizaki.2005.2007 (Temperature correction)
\bibitem{Ishizaki.2005.2007} A.~Ishizaki and Y.~Tanimura, J.~Phys.~Soc.~Jpn.
\textbf{74}, 3131 (2005); J.~Chem.~Phys. \textbf{125}, 084501 (2006);
J.~Phys.~Chem.~A \textbf{111}, 9269 (2007). 


%:Tanimura2006 (Review)
\bibitem{Tanimura2006} Y.~Tanimura, J.~Phys.~Soc.~Jpn. \textbf{75},
082001 (2006). 


%:YJ_Yan
\bibitem{YJ_Yan} R.-X.~Xu, P.~Cui, X.~Q.~Li, Y.~Mo, and Y.~J.~Yan,
J.~Chem.~Phys. \textbf{122}, 041103 (2004); R.-X.~Xu and Y.~J.~Yan,
Phys.~Rev.~E \textbf{75}, 031107 (2007); J.~Xu, R.-X.~Xu, and
Y.~J.~Yan, New J. Phys. \textbf{11}, 105037 (2009); L.~Chen, R.~Zheng,
Q.~Shi, and Y.~J.~Yan, J.~Chem.~Phys. \textbf{131}, 094502 (2009).


%:Ishizaki_FMO
\bibitem{Ishizaki_FMO} A.~Ishizaki and G.~R.~Fleming, PNAS \textbf{106},
17255 (2009); M.~Sarovar, A.~Ishizaki, G.~R.~Fleming, and K.~B.~Whaley,
Nature Phys. \textbf{6}, 462 (2010); J.~Strümpfer and K.~Schulten,
J.~Chem.~Phys. \textbf{131}, 225101 (2009); J.~Chem.~Phys. \textbf{134},
095102 (2011). %:JQ_You
 

\bibitem{JQ_You} J.~Q.~You and F.~Nori, Phys.~Today \textbf{58},
42 (2005); Nature \textbf{474}, 589 (2011). %:Tanimura2010PRL
 

\bibitem{Tanimura2010PRL} A.~G.~Dijkstra and Y.~Tanimura, Phys.~Rev.~Lett.
\textbf{104}, 250401 (2010). %:StrongCoupling_QED
 

\bibitem{StrongCoupling_QED} A.~Wallraff, \textit{et al}. Nature
\textbf{431}, 162 (2004); T.~Niemczyk, \textit{et al}. Nature Phys.
\textbf{6}, 772 (2010); 

%:DFS
\bibitem{DFS} P.~Zanardi and M.~Rasetti, Phys.~Rev.~Lett. \textbf{79},
3306 (1997); P.~Zanardi, Phys.~Rev.~A \textbf{56} 4445 (1997);
D.~A.~Lidar, I.~L.~Chuang, and K.~B.~Whaley, Phys.~Rev.~Lett.
\textbf{81}, 2594 (1998). 

%:t_yu (Ting Yu)
\bibitem{t_yu} T.~Yu and J.~H.~Eberly, Phys. Rev. Lett. \textbf{93},
140404 (2004); T.~Yu and J.~H.~Eberly, Science \textbf{323}, 598
(2009). 

%:Maniscalco2008PRL
\bibitem{Maniscalco2008PRL} S.~Maniscalco, F.~Francica, R.~L.~Zaffino,
N.~Lo Gullo, and F.~Plastina, Phys.~Rev.~Lett. \textbf{100}, 090503
(2008). 

%:Lamb shift
\bibitem{Hoeppe2012} U.~Hoeppe, C.~Wolff, J.~K\"uchenmeister, J.~Niegemann, M.~Drescher, H.~Benner, and K.~Busch, Phys. Rev. Lett. \textbf{108}, 043603 (2012).

%:Francia2009PRA 
\bibitem{Francia2009PRA} F.~Francica, S.~Maniscalco, J.~Piilo,
F.~Plastina, and K.-A.~Suominen, Phys.~Rev.~A \textbf{79}, 032310
(2009). 

%:Mazzola2009PRA
\bibitem{Mazzola2009PRA} L.~Mazzola, S.~Maniscalco, J.~Piilo,
K.-A.~Suominen, and B.~M.~Garraway, Phys.~Rev.~A \textbf{79},
042302 (2009). 

%:Non-RWA
\bibitem{Seke1993} J.~Seke, Physica A \textbf{193}, 587 (1993).

\bibitem{Intravaia2003} F.~Intravaia, S.~Maniscalco, and A.~Messina,
Eur. Phys. J. B \textbf{32}, 97 (2003).

\bibitem{Nesi2007} F.~Nesi, M.~Grifoni, and E.~Paladino, New J. Phys. \textbf{9}, 316 (2007).

\bibitem{Werlang2008} T. Werlang, A. V. Dodonov, E. I. Duzzioni, and C. J. Villas-B{\^o}as, Phys. Rev. A \textbf{78}, 053805 (2008).

\bibitem{Wang2008} F.-Q.~Wang, Z.-M. Zhang, and R.-S. Liang, Phys. Rev. A,
\textbf{78}, 062318 (2008).

\bibitem{XF_Cao} 
X.~Cao, and H.~Zheng, Phys. Rev. A \textbf{77}, 022320 (2008);
X.~Cao, and H.~Zheng, Eur.~Phys.~J.~B \textbf{68},
209 (2009). X.~Cao, J.~Q.~You, H.~Zheng, A.~G.~Kofman, and F.~Nori,
Phys.~Rev.~A \textbf{82}, 022119 (2010). 

\bibitem{JingJun} J.~Jing, Z.-G.~Lu, and Z.~Ficek, Phys.~Rev.~A
\textbf{79}, 044305 (2009); Z.~Ficek, J.~Jing, and Z.-G.~Lü, Phys.~Scr.
\textbf{T140}, 014005 (2010). 

\bibitem{Benatti} F.~Benatti, R.~Floreanini, and U.~Marzolino,
EPL \textbf{88}, 20011 (2009); F.~Benatti, R.~Floreanini, and U.~Marzolino,
Phys.~Rev.~A \textbf{81}, 012105 (2010). 

\bibitem{Altintas2012} F.~Altintas, and R.~Eryigit, Phys. Lett. A \textbf{376}, 1791 (2012).

%\bibitem{Garg1985}
 %A.~Garg, J.~N.~Onuchic, and V.~Ambegaokar, J.~Chem.~Phys. {\bf 83} 4491 (1985).
 \end{thebibliography}
\end{document}